\documentclass[]{pasj01}
\draft

\begin{document} 
\Received{}
\Accepted{}

\title{CO Multi-line Imaging of Nearby Galaxies (COMING):
I. Physical properties of molecular gas in the barred spiral galaxy NGC~2903}

 \author{%
   Kazuyuki \textsc{Muraoka}\altaffilmark{1},
   Kazuo \textsc{Sorai}\altaffilmark{2,3},
   Nario \textsc{Kuno}\altaffilmark{4,5},
   Naomasa \textsc{Nakai}\altaffilmark{4,5},
   Hiroyuki \textsc{Nakanishi}\altaffilmark{6},
   Miho \textsc{Takeda}\altaffilmark{1},
   Kazuki \textsc{Yanagitani}\altaffilmark{1},
   Hiroyuki \textsc{Kaneko}\altaffilmark{7},
   Yusuke \textsc{Miyamoto}\altaffilmark{7},
   Nozomi \textsc{Kishida}\altaffilmark{3},
   Takuya \textsc{Hatakeyama}\altaffilmark{4},
   Michiko \textsc{Umei}\altaffilmark{3},
   Takahiro \textsc{Tanaka}\altaffilmark{4},
   Yuto \textsc{Tomiyasu}\altaffilmark{4},   
   Chey \textsc{Saita}\altaffilmark{6},
   Saeko \textsc{Ueno}\altaffilmark{6},
   Naoko \textsc{Matsumoto}\altaffilmark{8,9},
   Dragan \textsc{SALAK}\altaffilmark{10},
   and 
   Kana \textsc{Morokuma-Matsui}\altaffilmark{9}
}

 \altaffiltext{1}{Department of Physical Science, Osaka Prefecture University, Gakuen 1-1, Sakai, Osaka 599-8531}
 \email{kmuraoka@p.s.osakafu-u.ac.jp}
 \altaffiltext{2}{Department of Physics, Faculty of Science, Hokkaido University, Kita 10 Nishi 8, Kita-ku, Sapporo 060-0810}
 \altaffiltext{3}{Department of Cosmosciences, Graduate School of Science, Hokkaido University, Kita 10 Nishi 8, Kita-ku, Sapporo 060-0810}
 \altaffiltext{4}{Division of Physics, Faculty of Pure and Applied Sciences, University of Tsukuba, 1-1-1 Tennodai, Tsukuba, Ibaraki 305-8571}
 \altaffiltext{5}{Center for Integrated Research in Fundamental Science and Technology (CiRfSE), University of Tsukuba, Tsukuba, Ibaraki 305-8571}
 \altaffiltext{6}{Graduate School of Science and Engineering, Kagoshima University, 1-21-35 Korimoto, Kagoshima, Kagoshima 890-0065}
 \altaffiltext{7}{Nobeyama Radio Observatory, Minamimaki, Minamisaku, Nagano 384-1305}
 \altaffiltext{8}{The Research Institute for Time Studies, Yamaguchi University, Yoshida 1677-1, Yamaguchi, Yamaguchi 753-8511}
 \altaffiltext{9}{National Astronomical Observatory of Japan, 2-21-1 Osawa, Mitaka, Tokyo 181-8588}
 \altaffiltext{10}{Department of Physics, School of Science and Technology, Kwansei Gakuin University, Gakuen 2-1, Sanda, Hyogo 669-1337}


\maketitle

\begin{abstract}
We present simultaneous mappings of $J=1-0$ emission of $^{12}$CO, $^{13}$CO, and C$^{18}$O molecules
toward the whole disk ($8' \times 5'$ or 20.8 kpc $\times$ 13.0 kpc) of the nearby barred spiral galaxy NGC~2903
with the Nobeyama Radio Observatory 45-m telescope at an effective angular resolution of $20''$ (or 870 pc).
We detected $^{12}$CO($J=1-0$) emission over the disk of NGC~2903.
In addition, significant $^{13}$CO($J=1-0$) emission was found at the center and bar-ends, whereas we could not detect any significant C$^{18}$O($J=1-0$) emission.
In order to improve the signal-to-noise ratio of CO emission and to obtain accurate line ratios
of $^{12}$CO($J=2-1$)/$^{12}$CO($J=1-0$) ($R_{2-1/1-0}$) and $^{13}$CO($J=1-0$)/$^{12}$CO($J=1-0$) ($R_{13/12}$),
we performed the stacking analysis for our $^{12}$CO($J=1-0$), $^{13}$CO($J=1-0$), and archival $^{12}$CO($J=2-1$) spectra with velocity-axis alignment in nine representative regions of NGC~2903.
We successfully obtained the stacked spectra of the three CO lines, and could measure averaged $R_{2-1/1-0}$ and $R_{13/12}$ with high significance for all the regions.
We found that both $R_{2-1/1-0}$ and $R_{13/12}$ differ according to the regions, which reflects the difference in the physical properties of molecular gas;
i.e., density ($n_{\rm H_2}$) and kinetic temperature ($T_{\rm K}$).
We determined $n_{\rm H_2}$ and $T_{\rm K}$ using $R_{2-1/1-0}$ and $R_{13/12}$ based on the large velocity gradient approximation.
The derived $n_{\rm H_2}$ ranges from $\sim 1000$ cm$^{-3}$ (in the bar, bar-ends, and spiral arms) to 3700 cm$^{-3}$ (at the center)
and the derived $T_{\rm K}$ ranges from 10 K (in the bar and spiral arms) to 30 K (at the center).
We examined the dependence of star formation efficiencies (SFEs) on $n_{\rm H_2}$ and $T_{\rm K}$,
and found the positive correlation between SFE and $n_{\rm H_2}$ with the correlation coefficient for the least-square power-law fit $R^2$ of 0.50.
This suggests that molecular gas density governs the spatial variations in SFEs.
\end{abstract}

\section{Introduction}

Molecular gas is one of the essential components in galaxies because it is closely related to star formation, which is a fundamental process of galaxy evolution.
Thus the observational study of molecular gas is indispensable to understand both star formation in galaxies and galaxy evolution.
However, the most abundant constituent in molecular gas, H$_2$, cannot emit any electro-magnetic wave
in cold molecular gas with typical temperature of $\sim$ 10 K due to the lack of a permanent dipole moment.
Instead, rotational transition lines of $^{12}$CO, the second abundant molecule, have been used as a tracer of molecular gas.
For example, some extensive $^{12}$CO surveys of external galaxies, which consist of single pointings toward central regions and some mappings along the major axis,
have been reported (e.g., \cite{braine1993}; \cite{young1995}; \cite{elfhag1996}).
These studies provided new findings about global properties of galaxies, such as excitation condition of molecular gas in galaxy centers
and radial distributions of molecular gas across galaxy disks.

In order to understand the relationship between molecular gas and star formation in galaxies further,
spatially resolved $^{12}$CO maps covering whole galaxy disks are necessary because star formation rates (SFRs) are often different between galaxy centers and disks.
In particular, single-dish observations are essential to measure $total$ molecular gas content in the observing beam from dense component to diffuse one avoiding the missing flux (e.g., \cite{caldu-primo2015}).
So far, two major surveys of wide-area $^{12}$CO mapping toward nearby galaxies are performed using multi-beam receivers mounted on large single-dish telescopes.

One is the $^{12}$CO($J=1-0$) mapping survey of 40 nearby spiral galaxies performed with the Nobeyama Radio Observatory (NRO) 45-m telescope in the position-switch mode (\cite{kuno2007}, hereafter K07).
Their $^{12}$CO($J=1-0$) maps cover most of the optical disks of galaxies at an angular resolution of 15$''$, and clearly show two-dimensional distributions of molecular gas in galaxies.
K07 found that the degree of the central concentration of molecular gas is higher in barred spiral galaxies than in non-barred spiral galaxies.
In addition, they found a correlation between the degree of central concentration and the bar strength adopted from \citet{Laurikainen2002}
\footnote{\citet{Laurikainen2002} estimated the maxima of the tangential force and the averaged radial force at each radius in a bar using {\it JHK} band images,
and they defined the maximum of the ratio between the two forces as the bar strength.};
i.e., galaxies with stronger bar tend to exhibit a higher central concentration.
This correlation suggests that stronger bars accumulate molecular gas toward central regions more efficiently,
which may contribute the onset of intense star formation at galaxy centers (i.e., higher SFRs than disks).
Using the $^{12}$CO($J=1-0$) data, \citet{sorai2012} investigated the physical properties of molecular gas in the barred spiral galaxy Maffei 2.
They found that molecular gas in the bar ridge regions may be gravitationally unbound, which suggests that molecular gas is hard to become dense, and to form stars in the bar.

The other survey is the Heterodyne Receiver Array CO Line Extragalactic Survey performed with the IRAM 30-m telescope \citep{leroy2009}.
They observed $^{12}$CO($J=2-1$) emission over the full optical disks of 48 nearby galaxies at an angular resolution of 13$''$,
and found that the $^{12}$CO($J=2-1$)/$^{12}$CO($J=1-0$) line intensity ratio (hereafter $R_{2-1/1-0}$) typically ranges from 0.6 to 1.0 with the averaged value of 0.8.
In addition, \citet{leroy2013} examined a quantitative relationship between surface densities of molecular gas and SFRs for 30 nearby galaxies at a spatial resolution of 1~kpc using the $^{12}$CO($J=2-1$) data.
They found a first-order linear correspondence between surface densities of molecular gas and SFRs but also found second-order systematic variations;
i.e., the apparent molecular gas depletion time, which is defined by the ratio of the surface density of molecular gas to that of SFR,
becomes shorter with the decrease in stellar mass, metallicity, and dust-to-gas ratio. They suggest that this can be explained by a CO-to-H$_2$ conversion factor ($X_{\rm CO}$) that depends on dust shielding.

However, such global CO maps of galaxies have raised a new question; the cause of the spatial variation in star formation efficiencies (SFEs) defined as SFRs per unit gas mass
\footnote{In this paper, a correction factor to account for Helium and other heavy elements is not included in the calculation of molecular gas mass and SFE.}.
It is reported that SFEs differ not only among galaxies (e.g., \cite{young1996}) but also within locations/regions in a galaxy (e.g., \cite{muraoka2007});
i.e., higher SFEs are often observed in galaxy mergers rather than normal spiral galaxies and also observed in the nuclear star forming region rather than in galaxy disks.
Some observational studies based on HCN emission, an excellent dense gas tracer, suggest that
SFEs increase with the increase in molecular gas density (or dense gas fraction) in galaxies (e.g., \cite{gao2004}; \cite{gao2007}; \cite{muraoka2009}; \cite{usero2015}),
but the cause of the spatial variation in SFEs is still an open question because HCN emission in galaxy disks is too weak to obtain its map except for some gas-rich spiral galaxies (e.g., M~51; \cite{chen2015}; \cite{bigiel2016}).

Instead, isotopes of CO molecule are promising probes of molecular gas density.
In particular, $^{13}$CO($J=1-0$) is thought to be optically thin and thus trace denser molecular gas ($\sim 10^{3-4} {\rm cm}^{-3}$)
rather than $^{12}$CO($J=1-0$), which is optically thick and traces relatively diffuse molecular gas ($\sim 10^{2-3} {\rm cm}^{-3}$).
Therefore, the relative intensity between $^{13}$CO($J=1-0$) and $^{12}$CO($J=1-0$) is  sensitive to physical properties of molecular gas.
For example, spatial variations in $^{13}$CO($J=1-0$)/$^{12}$CO($J=1-0$) intensity ratios (hereafter $R_{13/12}$) were observed in nearby galaxy disks (e.g., \cite{sakamoto1997}; \cite{tosaki2002}; \cite{hirota2010}).
Such variations in $R_{13/12}$, typically ranging from 0.05 to 0.20, are interpreted as the variation in molecular gas density; i.e., $R_{13/12}$ increases with the increase in molecular gas density.
However, some observations suggest that $R_{13/12}$ in central regions of nearby galaxies are lower than those in disk regions (e.g., \cite{paglione2001}; \cite{tosaki2002}; \cite{hirota2010}; \cite{watanabe2011})
although central regions of galaxies often show intense star formation activities, suggesting higher molecular gas density.
The cause of the low $R_{13/12}$ in central regions is thought to be high temperature of molecular gas due to the heating by UV radiation from a lot of young massive stars.

Such a degeneracy between density and temperature of molecular gas in a line ratio can be solved using two (or more) molecular line ratios
with a theoretical calculation on the excitation of molecular gas such as the large velocity gradient (LVG) model \citep{scoville1974, goldreich1974}.
For example, the density and kinetic temperature of giant molecular clouds (GMCs) were determined using $R_{13/12}$ and $^{12}$CO($J=3-2$)/$^{12}$CO($J=1-0$) ratio
for Large Magellanic Cloud \citep{minamidani2008} and M~33 \citep{muraoka2012}, and also determined using $R_{13/12}$ and $R_{2-1/1-0}$ for the spiral arm of M~51 \citep{schinnerer2010}.
This method to determine molecular gas density is useful to investigate the cause of the variation in SFEs.
Thus the dependence of SFEs on molecular gas density should be investigated for various galaxies at high angular resolution based on multiple line ratios including $R_{13/12}$.

In this paper, we investigate the relationship between SFE and molecular gas density within a nearby barred spiral galaxy NGC~2903
using an archival $^{12}$CO($J=2-1$) map combined with $^{12}$CO($J=1-0$) and $^{13}$CO($J=1-0$) maps which are newly obtained
by the CO Multi-line Imaging of Nearby Galaxies (COMING) project with the NRO 45-m telescope.
NGC~2903 is a gas-rich galaxy exhibiting bright nuclear star formation (e.g., \cite{wynn1985}; \cite{simons1988}; \cite{alonso2001}; \cite{yukita2012}).
The distance to NGC~2903 is estimated to be 8.9 Mpc \citep{drozdovsky2000}; thus the effective angular resolution of 20$''$ for the on-the-fly (OTF) mapping with the NRO 45-m corresponds to 870 pc.
This enables us to resolve major structures within NGC~2903, such as the center, bar, and spiral arms
although its inclination of \timeform{65D} \citep{deblok2008} is not so small.
In addition, NGC~2903 is rich in archival multi-wavelength data set; i.e., not only the $^{12}$CO($J=2-1$) map to examine $R_{2-1/1-0}$ but also H$\alpha$ and infrared images to calculate SFRs are available.
Thus this galaxy is a preferable target to examine the cause of the variation in SFE in terms of molecular gas density.
Basic parameters of NGC~2903 are summarized in table~1.

The structure of this paper is as follows:
We describe the overview of the COMING project and explain the detail of the CO observations and data reduction for NGC~2903 in section 2.
Then, we show results of observations; i.e., spectra and velocity-integrated intensity maps of $^{12}$CO($J=1-0$) and $^{13}$CO($J=1-0$) emission in section 3.
We obtain averaged spectra of $^{12}$CO($J=1-0$), $^{13}$CO($J=1-0$), and $^{12}$CO($J=2-1$) emission for nine representative regions,
and measure averaged $R_{13/12}$ and $R_{2-1/1-0}$ for each region in section 4.1.
We determine molecular gas density and kinetic temperature for the center, bar, bar-ends, and spiral arms using $R_{13/12}$ and $R_{2-1/1-0}$ based on the LVG approximation in section 4.2.
Finally, we investigate the cause of the variation in SFE by examining the dependence of SFE on molecular gas density and kinetic temperature.

\section{Observations and data reduction}

\subsection{COMING project}

COMING is a project to map $J=1-0$ emission of $^{12}$CO, $^{13}$CO, and C$^{18}$O molecules simultaneously for 70\% area of optical disks of 238 galaxies
using the FOur-beam REceiver System on 45-m Telescope (FOREST; \cite{minamidani2016}) at NRO.
The main purposes of the COMING are to characterize properties of molecular gas as sequence of, Hubble types, dynamical structures,
central concentrations, and star formation activities, as well as surrounding environments of galaxies.
More detailed information on COMING project including the current status of the survey will be reported in the forthcoming paper (Sorai et al. in preparation).

\subsection{Observations}

$^{12}$CO($J=1-0$), $^{13}$CO($J=1-0$), and C$^{18}$O($J=1-0$) emission observations of NGC~2903 were performed
using the NRO 45-m telescope from April to May, 2015, employing the OTF mapping mode.
The observed area is about $8' \times 5'$, which corresponds to $20.8 \times 13.0$ kpc at the distance of 8.9 Mpc, as indicated in figure~1.
The total time for the observations was 13 hrs.

We used a new $2 \times 2$ focal-plane dual-polarization sideband-separating SIS mixer receiver for the single side band (SSB) operation, FOREST,
which provides 8 intermediate frequency (IF) paths (i.e., 4 beam $\times$ 2 polarization) independently.
Owing to the wide IF range of 4 to 12 GHz, we could simultaneously observe $^{12}$CO($J=1-0$) emission at 115 GHz (IF = 10 GHz) and
$^{13}$CO($J=1-0$) and C$^{18}$O($J=1-0$) emission at 110 GHz (IF = 5 GHz) when the frequency of the local oscillator was set to 105 GHz.
The backend was an FX-type correlator system, SAM45, which consists of 16 arrays with 4096 spectral channels each.
We set the frequency coverage and resolution for each array of 2 GHz and 488.24 kHz,
which gives velocity coverage and resolution of 5220 km s$^{-1}$ and 1.27 km s$^{-1}$ at 115 GHz, and those of 5450 km s$^{-1}$ and 1.33 km s$^{-1}$ at 110 GHz.
We assigned 8 of 16 arrays to 115 GHz band (i.e., IF = 9 -- 11 GHz for $^{12}$CO($J=1-0$) emission)
and other 8 arrays to 110 GHz band (i.e., IF = 4 -- 6 GHz for $^{13}$CO($J=1-0$) and C$^{18}$O($J=1-0$) emission).
The half-power beam widths of the 45-m with the FOREST were $\sim 14''$ at 115 GHz and $\sim 15''$ at 110 GHz, respectively.
The system noise temperatures were 300 -- 500 K at 115 GHz and 200 -- 250 K at 110 GHz during the observing run.

We performed the OTF mapping along the major and minor axes of the galaxy disk whose position angle was \timeform{25D} (K07).
The separation between the scan rows was set to $5\farcs0$, and the spatial sampling interval was $<2\farcs4$ applying a dump time of 0.1~second and a scanning speed of $<{24''}$~s$^{-1}$.
The data sets scanned along two orthogonal axes were co-added by the Basket-weave method \citep{emerson1988} to remove any effects of scanning noise.
In order to check the absolute pointing accuracy every hour, we observed an SiO maser source, W-Cnc, using a 43 GHz band receiver.
It was better than $6''$ (peak-to-peak) throughout the observations.
In addition, we observed $^{12}$CO($J=1-0$) and $^{13}$CO($J=1-0$) emission of W~3 and IRC+10216 every day
to obtain the scaling factors for converting the observed antenna temperature to the main beam temperature for each IF.
Note that these scaling factors not only correct the main-beam efficiency ($\eta_{\rm MB}$) of the 45-m antenna
but also compensate the decrease in line intensity due to the incompleteness of the image rejection for SSB receiver (e.g., \cite{nakajima2013}).
The absolute error of the temperature scale for each CO spectrum was about $\pm20$\%, mainly due to variations in $\eta_{\rm MB}$ and the image rejection ratio of the FOREST.

\subsection{Data reduction}

Data reduction was made using the software package NOSTAR, which comprises tools for OTF data analysis, developed by NRO \citep{sawada2008}. 
We excluded bad spectra, which includes strong baseline undulation and spurious lines, from the raw data.
Then, linear baselines were subtracted, and the raw data were regridded to 6$^{\prime \prime}$ per pixel
with an effective angular resolution of approximately 20$^{\prime \prime}$ (or 870 pc).
We binned the adjacent spectral channels to a velocity resolution of 10 km s$^{-1}$ for the spectra.
Finally, we created three-dimensional data cubes in $^{12}$CO($J=1-0$), $^{13}$CO($J=1-0$), and C$^{18}$O($J=1-0$) emission.
The resultant r.m.s.\ noise levels (1 $\sigma$) were 60 mK, 39 mK, and 40 mK for $^{12}$CO($J=1-0$), $^{13}$CO($J=1-0$), and C$^{18}$O($J=1-0$), respectively.

\section{Results}

\subsection{$^{12}$CO($J=1-0$) emission}

Figure~2 shows $^{12}$CO($J=1-0$) spectra of the whole optical disk in NGC~2903.
As is the case in earlier studies (e.g., \cite{helfer2003}, K07, \cite{leroy2009}),
strong $^{12}$CO($J=1-0$) emission, whose peak temperature was $\sim$ 0.6 K, was found at the center
and significant $^{12}$CO($J=1-0$) emission was detected in the bar ($\sim$ 0.3 K), bar-ends (0.4 -- 0.5 K), and spiral arms (0.1 -- 0.3 K).

We calculate velocity-integrated $^{12}$CO($J=1-0$) intensities ($I_{\rm 12CO(1-0)}$) from the spectra.
In order to obtain more accurate line intensities (in other words, to minimalize the effects of the noise and the undulation of baseline for weak line),
we defined the ``line channels'', which are successive velocity channels where significant emission exists, in advance for each pixel as described below.

In order to define the ``line channels'', we utilized $^{12}$CO($J=2-1$) data (\cite{leroy2009}),
which was regridded and convolved to 20$^{\prime \prime}$ to match our $^{12}$CO($J=1-0$) spectra.
Since the 1 $\sigma$ r.m.s.\ of $^{12}$CO($J=2-1$) data of 6 mK at 20$^{\prime \prime}$ and 10 km s$^{-1}$ resolutions was 10 times better than that of our $^{12}$CO($J=1-0$) data,
the $^{12}$CO($J=2-1$) spectra are appropriate for the decision of ``line channels'' in each pixel.
We first identified a velocity channel exhibiting the peak $^{12}$CO($J=2-1$) temperature and defined the channel as the ``CO peak channel'' for each pixel.
Then, successive channels whose $^{12}$CO($J=2-1$) emission consistently exceeds 1 $\sigma$ including the ``CO peak channel'' are defined as ``line channels''.
Finally, we calculated $I_{\rm 12CO(1-0)}$ for the specified ``line channels'' in each pixel.

Figure~3 shows the $I_{\rm 12CO(1-0)}$ map of NGC~2903.
The strongest $I_{\rm 12CO(1-0)}$ of 92 K km s$^{-1}$ is observed at the center, and the secondary peak of 55 K km s$^{-1}$ is at the northern bar-end.
The total molecular gas mass for the observed area in NGC~2903 is estimated to ($2.8\pm0.6$) $\times 10^9$ $M_{\odot}$
under the assumptions of the constant $X_{\rm CO}$ of $1.8 \times 10^{20}$ ${\rm cm}^{-2}$ (K km s$^{-1}$)$^{-1}$ \citep{dame2001} over the disk
and the uncertainty of 20\% in brightness temperature scape of CO line.
This value is consistent with $3.2 \times 10^9$ $M_{\odot}$ obtained by K07, which is recalculated using the same distance and $X_{\rm CO}$.
We also compare $I_{\rm 12CO(1-0)}$ obtained by COMING with those obtained by K07 to confirm the validity of our $^{12}$CO($J=1-0$) data.
We examined the pixel-by-pixel comparison for the two $I_{\rm 12CO(1-0)}$ maps at the same angular resolution of 20$''$ as shown in figure~4,
and confirmed that both $I_{\rm 12CO(1-0)}$ are well correlated with each other. 
The median and the standard deviation in $I_{\rm 12CO(1-0)}$ are 12.9 K km s$^{-1}$ and 12.4 K km s$^{-1}$ for COMING dataset,
and those are 15.2 K km s$^{-1}$ and 12.6 K km s$^{-1}$ for K07 dataset.

\subsection{$^{13}$CO($J=1-0$) and C$^{18}$O($J=1-0$) emission}

Figure~5(a) shows the global $^{13}$CO($J=1-0$) spectra, which are overlaid by $^{12}$CO($J=1-0$) spectra for comparison.
In addition, figure~5(b), (c), and (d) show the magnified $^{13}$CO($J=1-0$) spectra at the northern bar-end, the center, and the southern bar-end, respectively.
We found significant $^{13}$CO($J=1-0$) emission at the center and both bar-ends.
However, we could not detect any significant C$^{18}$O($J=1-0$) emission.

We calculated the velocity-integrated $^{13}$CO($J=1-0$) intensities ($I_{\rm 13CO(1-0)}$).
As is the case of $^{12}$CO($J=1-0$), we utilized the ``line channels'' defined by $^{12}$CO($J=2-1$) spectra.
Figure~6 shows a spatial distribution of $I_{\rm 13CO(1-0)}$ in pseudo-color overlaid by $I_{\rm 12CO(1-0)}$ in contour.
The global distribution of $^{13}$CO($J=1-0$) is similar to $^{12}$CO($J=1-0$);
several peaks whose $I_{\rm 13CO(1-0)}$ exceeds 5 K km s$^{-1}$ are observed at the center, bar-ends, and in spiral arms.

\subsection{Line intensity ratios $R_{2-1/1-0}$ and $R_{13/12}$}

Intensity ratios of two (or more) molecular lines provide important clues to estimate physical properties of molecular gas, such as density and temperature. 
We examined the spatial variations in line intensity ratios among $^{12}$CO($J=1-0$), $^{13}$CO($J=1-0$), and $^{12}$CO($J=2-1$) emission.
Figure~7 shows spatial distributions of $R_{2-1/1-0}$ and $R_{13/12}$ over the disk of NGC~2903.
In these maps, we displayed pixels with each line intensity exceeding 2 $\sigma$
(5 K km s$^{-1}$, 3  K km s$^{-1}$, and 0.5 K km s$^{-1}$ for $^{12}$CO($J=1-0$), $^{13}$CO($J=1-0$), and $^{12}$CO($J=2-1$), respectively).
We found some local peaks of $R_{2-1/1-0}$ ($\sim$ 1.0) near the center and at the downstream side of the northern spiral arm,
whereas lower $R_{2-1/1-0}$ ($\sim$ 0.5 -- 0.6) was observed in the bar.
The spatial distribution of $R_{13/12}$ is unclear due to poor signal-to-noise (S/N) ratio of $^{13}$CO($J=1-0$) emission.

\section{Discussion}

\subsection{Velocity-axis alignment stacking of CO spectra}

As described in section 3.3, the spatial distribution of $R_{13/12}$ seems noisy and unclear due to the poor S/N although we could obtain the spatial distribution of $I_{\rm 13CO(1-0)}$.
In order to improve the S/N of weak emission such as $^{13}$CO($J=1-0$), the stacking analysis of CO spectra with the velocity-axis alignment seems a promising method.

The stacking technique for CO spectra in external galaxies are originally demonstrated by Schruba et al. (2011, 2012).
Since the observed velocities of each position within a galaxy are different due to its rotation, a simple stacking causes a smearing of spectrum.
In order to overcome such difficulty, \citet{schruba2011} demonstrated the velocity-axis alignment of CO spectra
in different regions in a galaxy disk according to mean H\emissiontype{I} velocity.
They stacked velocity-axis aligned CO spectra, and successfully confirmed very weak $^{12}$CO($J=2-1$) emission ($<$ 1 K km s$^{-1}$)
with high significance in H\emissiontype{I}-dominated outer-disk regions of nearby spiral galaxies.
In addition, \citet{schruba2012} applied this stacking technique to perform the sensitive search for weak $^{12}$CO($J=2-1$) emission in dwarf galaxies.
Furthermore, \citet{morokuma-matsui2015} applied the stacking technique to $^{13}$CO($J=1-0$) emission in the optical disk of the nearby barred spiral galaxy NGC~3627.
By the stacking with velocity-axis alignment based on mean $^{12}$CO($J=1-0$) velocity,
they obtained high S/N $^{13}$CO($J=1-0$) spectra which are improved
by a factor of up to 3.2 compared to the normal (without velocity-axis alignment) stacking analysis.

These earlier studies clearly suggest that the stacking analysis is very useful to detect weak molecular line.
In this section, we employ the same stacking technique as \citet{morokuma-matsui2015}
to improve the S/N of $^{13}$CO($J=1-0$) emission and to obtain more accurate line ratios.
Based on our $I_{\rm 12CO(1-0)}$ image (figure~3), we have separated NGC~2903 into nine regions according to its major structures;
i.e., center, northern bar, southern bar, northern bar-end, southern bar-end, northern arm, southern arm, inter-arm, and outer-disk.
The left panel of figure~8 shows the separation of each region overlaid by the grey-scale map of $I_{\rm 12CO(1-0)}$.
For each region, we stacked $^{12}$CO($J=1-0$), $^{13}$CO($J=1-0$), and $^{12}$CO($J=2-1$) spectra with velocity-axis alignment
based on the intensity-weighted mean velocity field calculated from our $^{12}$CO($J=1-0$) data (right panel of figure~8).
We successfully obtained the stacked CO spectra as shown in figure~9.
The S/N of each CO emission is dramatically improved, and thus we could confirm the significant $^{13}$CO($J=1-0$) emission for all the regions.
We found the difference in the line shape of stacked CO spectra according to regions.
In particular, the stacked $^{12}$CO spectra in the bar show flat peak over the velocity width of 100 -- 150 km s$^{-1}$.
This is presumably due to the rapid velocity change in the bar, which makes the velocity-axis alignment difficult.

We summarize the averaged line intensities and line ratios for each region in table~2.
We found that the averaged $R_{2-1/1-0}$ shows the highest value of 0.92 at the center,
and a moderate value of 0.7 -- 0.8 at both bar-ends and in the northern arm.
A slightly lower $R_{2-1/1-0}$ of 0.6 -- 0.7 is observed in the bar, southern arm, inter-arm, and outer-disk.
Such a variation in $R_{2-1/1-0}$ ranging from 0.6 to 1.0 in NGC~2903 is quite consistent with those observed in nearby galaxies (e.g., \cite{leroy2009}).
However, the highest $R_{13/12}$ of 0.19 is observed not at the center but in the northern arm.
The central $R_{13/12}$ of 0.11 is similar to those in other regions (0.08 -- 0.13) 
except for the northern arm and outer-disk ($\sim$ 0.04).
The typical $R_{13/12}$ of $\sim$ 0.1 is frequently observed in nearby galaxies (e.g., \cite{paglione2001}; \cite{vila2015}),
but slightly higher than the averaged $R_{13/12}$ in representative regions of NGC~3627, 0.04 -- 0.09 \citep{morokuma-matsui2015}.

\subsection{Derivation of physical properties and their comparison with star formation}

\subsubsection{LVG calculation for stacked CO spectra}

Using $R_{2-1/1-0}$ and $R_{13/12}$, we derive averaged physical properties of molecular gas,
its density ($n_{\rm H_2}$) and kinetic temperature ($T_{\rm K}$), in seven regions (center, northern bar, southern bar, northern bar-end, southern bar-end, northern arm, and southern arm) of NGC~2903 based on the LVG approximation.
Some assumptions are required to perform the LVG calculation; the molecular abundances $Z$($^{12}$CO) = [$^{12}$CO]/[H$_2$],
[$^{12}$CO]/[$^{13}$CO], and the velocity gradient $dv/dr$.
Firstly, we fix the $Z$($^{12}$CO) of $1.0 \times 10^{-5}$ and $dv/dr$ of 1.0 km s$^{-1}$ pc$^{-1}$;
i.e., $^{12}$CO abundance per unit velocity gradient $Z$($^{12}$CO)/($dv/dr$) was assumed to $1.0 \times 10^{-5}$ (km s$^{-1}$ pc$^{-1}$)$^{-1}$.
This is the same as the assumed $Z$($^{12}$CO)/($dv/dr$) for GMCs in M~33 \citep{muraoka2012}.

Then, we determine the [$^{12}$CO]/[$^{13}$CO] abundance ratio to be assumed in this study by considering earlier studies.
\citet{langer1990} found a systematic gradient in the $^{12}$C/$^{13}$C isotopic ratio across in our Galaxy;
from $\sim$ 30 in the inner part at 5 kpc to $\sim$ 70 at 12 kpc with a galactic center value of 24.
For external galaxies, the reported $^{12}$C/$^{13}$C isotopic ratios in their central regions are 40 for NGC~253 \citep{henkel1993},
50 for NGC~4945 \citep{henkel1994}, $> 40$ for M~82 and $> 30$ for IC~342 \citep{henkel1998}.
\citet{mao2000} reported a higher [$^{12}$CO]/[$^{13}$CO] abundance ratio of 50 -- 75 in the central region of M~82.
\citet{martin2010} also reported a higher $^{12}$C/$^{13}$C isotopic ratios of $>$ 50 -- 100 in the central regions of M~82 and NGC~253.
In summary, reported $^{12}$C/$^{13}$C isotopic (and [$^{12}$CO]/[$^{13}$CO] abundance) ratios in nearby galaxy centers (30 -- 100)
are typically higher than that in the inner 5 kpc of our Galaxy (24 -- 30),
but the cause of such discrepancies in $^{12}$C/$^{13}$C and [$^{12}$CO]/[$^{13}$CO] between our Galaxy and external galaxies is still unresolved.
Here, we assumed an intermediate [$^{12}$CO]/[$^{13}$CO] abundance ratio of 50 in NGC~2903 without any gradient across its disk for our LVG calculation.
Note that we perform an additional LVG calculation for the center of NGC~2903 assuming the [$^{12}$CO]/[$^{13}$CO] abundance ratios of 30 and 70
to evaluate the effect of the variation in the assumed [$^{12}$CO]/[$^{13}$CO] abundance ratio on results of LVG calculation.

Figure~10 shows results of LVG calculation for each region in NGC~2903.
The thin line indicates a curve of constant $R_{2-1/1-0}$ as functions of $n_{\rm H_2}$ and $T_{\rm K}$, 
and the thick line indicates that of constant $R_{13/12}$.
We can determine both $n_{\rm H_2}$ and $T_{\rm K}$ at the point where two curves intersect each other.
Under the assumption of [$^{12}$CO]/[$^{13}$CO] abundance ratio of 50,
the derived $n_{\rm H_2}$ ranges from $\sim$1000 cm$^{-3}$ (in the disk; i.e., bar, bar-ends, and spiral arms) to 3700 cm$^{-3}$ (at the center)
and the derived $T_{\rm K}$ ranges from 10 K (in spiral arms) to 30 K (at the center).
Note that both $n_{\rm H_2}$ and $T_{\rm K}$ vary depending on the assumption of [$^{12}$CO]/[$^{13}$CO] abundance ratio;
at the center of NGC~2903, the abundance ratio of 30 yields lower $n_{\rm H_2}$ of 1800 cm$^{-3}$ and higher $T_{\rm K}$ of 38 K,
whereas the abundance ratio of 70 yields higher $n_{\rm H_2}$ of 5900 cm$^{-3}$ and intermediate $T_{\rm K}$ of 29 K.
It seems that $n_{\rm H_2}$ is proportional to [$^{12}$CO]/[$^{13}$CO] abundance ratio.
This trend of $n_{\rm H_2}$ can be naturally explained if we consider the optical depth of $^{12}$CO and $^{13}$CO emission.
$^{12}$CO is always optically thick and thus its emission emerges from the diffuse envelope of dense gas clouds,
while $^{13}$CO emission emerges from further within these dense gas clouds due to its lower abundance.
Since the increase in the assumed [$^{12}$CO]/[$^{13}$CO] abundance ratio means that $^{13}$CO becomes more optically thin,
$^{13}$CO emission emerged from deeper within the dense gas clouds and thus it probes denser gas.
Derived physical properties, $n_{\rm H_2}$ and $T_{\rm K}$, are summarized in table~3.

We compare the derived $n_{\rm H_2}$ and $T_{\rm K}$ in NGC~2903 with those determined in other external galaxies.
\citet{muraoka2012} determined $n_{\rm H_2}$ and $T_{\rm K}$ for GMCs associated with the giant H\emissiontype{II} region NGC~604 in M~33 at a spatial resolution of 100 pc
using three molecular lines, $^{12}$CO($J=1-0$), $^{13}$CO($J=1-0$), and $^{12}$CO($J=3-2$), based on the LVG approximation.
The derived $n_{\rm H_2}$ and $T_{\rm K}$ are 800 -- 2500 cm$^{-3}$ and 20 -- 30 K, respectively,
which are similar to our study for NGC~2903 in spite of the difference in the spatial resolution.

However, \citet{schinnerer2010} obtained different physical properties for GMCs in spiral arms of M~51.
They performed the LVG analysis using $R_{13/12}$ and $R_{2-1/1-0}$ at a spatial resolution of 120 -- 180 pc.
For the case of constant $dv/dr$ = 1.0 km s$^{-1}$ pc$^{-1}$, the derived $T_{\rm K}$ ranges from 10 to 50 K, which is similar to our study for NGC~2903,
whereas $n_{\rm H_2}$ ranges from 100 to 400 cm$^{-3}$, which is 5 -- 10 times lower than that in the disk of NGC~2903
in spite that the values of $R_{2-1/1-0}$ and $R_{13/12}$ in M~51 are not so different from those in NGC~2903.
This is presumably due to the differences in assumed $Z$($^{12}$CO) and [$^{12}$CO]/[$^{13}$CO] abundance ratio.
The authors assumed $Z$($^{12}$CO) of 8.0 $\times 10^{-5}$, which is higher than that assumed in our study, and a lower [$^{12}$CO]/[$^{13}$CO] abundance ratio of 30.
Under the LVG approximation with the assumption of $Z$($^{12}$CO) of 8.0 $\times 10^{-5}$,
we found that the derived $n_{\rm H_2}$ is typically $\sim$3 times lower than that with assumption of $Z$($^{12}$CO) of 1.0 $\times 10^{-5}$.
Physically, high $Z$($^{12}$CO) means abundant $^{12}$CO molecules among molecular gas.
In this condition, the optical depth of $^{12}$CO line also increases, and thus the photon-trapping effect in molecular clouds becomes effective.
Since this effect contributes the excitation of $^{12}$CO molecule, an effective critical density of $^{12}$CO line decreases.
In other words, since the $^{12}$CO is easily excited to upper $J$ levels even in low molecular gas density, $n_{\rm H_2}$ at a given $R_{2-1/1-0}$ decreases.
As a result, LVG analysis with the assumption of $Z$($^{12}$CO) of 8.0 $\times 10^{-5}$ yields lower $n_{\rm H_2}$.
In addition, the low [$^{12}$CO]/[$^{13}$CO] abundance ratio of 30 causes the decrease in the derived molecular gas density as described above.
Therefore, the difference in the derived $n_{\rm H_2}$ between NGC~2903 and M~51 can be explained by the difference in the assumed $Z$($^{12}$CO) and the [$^{12}$CO]/[$^{13}$CO] abundance ratio.

\subsubsection{Comparison of SFE with density and kinetic temperature of molecular gas}

As described in section 1, SFEs often differ between galaxy centers and disks.
Since NGC~2903 has a bright star forming region at the center, its SFE is expected to be higher than those in other regions.
Here, we calculate SFEs for seven regions where averaged physical properties of molecular gas are obtained,
and compare SFE with $n_{\rm H_2}$ and $T_{\rm K}$ in each region to examine what parameter controls SFE in galaxies.

SFE is expressed using the surface density of SFR ($\Sigma_{\rm SFR}$) and that of molecular hydrogen ($\Sigma_{\rm H_2}$) as follows:
\begin{eqnarray}
\left[ \frac{\rm SFE}{\rm yr^{-1}} \right]= \left( \frac{\Sigma_{\rm SFR}}{M_{\odot}\,{\rm yr^{-1}\,pc^{-2}}} \right) {\displaystyle \biggl/} \left( \frac{\Sigma_{\rm H_2}}{M_{\odot}\,{\rm pc^{-2}}} \right) 
\end{eqnarray}
We calculated extinction-corrected SFRs from a linear combination of H$\alpha$ and $Spitzer$/MIPS 24 $\micron$ luminosities
using a following formula \citep{kennicutt1998a, kennicutt1998b, calzetti2007}:
\begin{eqnarray}
{\Sigma_{\rm SFR}} = 7.9 \times 10^{-42} \left( \frac{L_{{\rm H} \alpha } + 0.031 \times L_{24 \mu {\rm m}}}{{\rm erg} \,\, {\rm s}^{-1}} \right) \frac{{\rm cos} \ i}{\Omega} \,\,\,\,\,\,\,\, M_{\odot} \,\, {\rm yr}^{-1} \,\, {\rm pc}^{-2},
\end{eqnarray}
where $L_{{\rm H} \alpha}$ and $L_{24 \mu {\rm m}}$ mean H$\alpha$ and 24 $\micron$ luminosities, respectively.
$i$ is the inclination of \timeform{65D} for NGC~2903 and $\Omega$ is the covered area for each region (in the unit of pc$^{2}$).
We used archival continuum-subtracted H$\alpha$ and 24 $\micron$ images of NGC~2903 obtained by \citet{hoopes2001} and the Local Volume Legacy survey project \citep{kennicutt2008, dale2009}, respectively.
In addition, we calculated $\Sigma_{\rm H_2}$ using $I_{\rm 12CO(1-0)}$ as follows:
\begin{eqnarray}
\left[ \frac{\Sigma_{\rm H_2}}{M_{\odot}\,{\rm pc^{-2}}} \right] &=& 2.89 \times {\rm cos} \ i \left( \frac{I_{\rm 12CO(1-0)}}{{\rm K\,\,km\,\,s^{-1}}} \right) \times \left\{ \frac{X_{\rm CO}}{1.8 \times 10^{20}\,{\rm cm}^{-2}\,({\rm K\,\,km\,\,s^{-1}})^{-1}} \right\} .
\end{eqnarray}
Here, we adopted a constant $X_{\rm CO}$ value of $1.8 \times 10^{20}$ ${\rm cm}^{-2}$ (K km s$^{-1}$)$^{-1}$ \citep{dame2001}.
We found that SFE at the center, $6.8 \times 10^{-9}$ yr$^{-1}$, is 2 -- 4 times higher than those in other regions.
Calculated SFEs are listed in table~3.

We examined the dependence of SFE on $n_{\rm H_2}$ and $T_{\rm K}$ as shown in figure~11. We found that SFE positively correlates with both $n_{\rm H_2}$ and $T_{\rm K}$.
However, the trend of these correlations might change because it is possible that variations in the [$^{12}$CO]/[$^{13}$CO] abundance ratio and $X_{\rm CO}$ affect the estimate of $n_{\rm H_2}$, $T_{\rm K}$, and SFE.
In fact, both the [$^{12}$CO]/[$^{13}$CO] abundance ratio and $X_{\rm CO}$ often differ between galaxy centers and disks.
Therefore, we examine how variations in the [$^{12}$CO]/[$^{13}$CO] abundance ratio and $X_{\rm CO}$ alter the estimate of $n_{\rm H_2}$, $T_{\rm K}$, and SFE at the center of NGC~2903.

We first consider the effect of the variation in [$^{12}$CO]/[$^{13}$CO] abundance ratio on the estimate of $n_{\rm H_2}$ and $T_{\rm K}$.
As described in section 4.2.1, it is reported that the $^{12}$C/$^{13}$C abundance ratio in our Galaxy increases with the galactocentric radius \citep{langer1990}.
Thus we examine the case of the lower [$^{12}$CO]/[$^{13}$CO] abundance ratio at the center of NGC~2903.
If we adopt the [$^{12}$CO]/[$^{13}$CO] abundance ratio of 30 at the center, $n_{\rm H_2}$ and $T_{\rm K}$ are estimated to be 1800 cm$^{-3}$ and 38 K, respectively.
This $n_{\rm H_2}$ value is slightly lower than that in the northern arm, whereas the positive correlation between SFE and $n_{\rm H_2}$ is not destroyed.
Similarly, the $T_{\rm K}$ of 38 K does not destroy the positive correlation between SFE and $T_{\rm K}$.

Next, we consider the effect of the variation in $X_{\rm CO}$ on the estimate of SFE.
In central regions of disk galaxies, $X_{\rm CO}$ drops (i.e., CO emission becomes luminous at a given gas mass) by a factor of 2 -- 3 or more
(e.g., \cite{nakai1995}; \cite{regan2000}), including the Galactic Center (e.g., \cite{oka1998}; \cite{dahmen1998}).
Such a trend is presumably applicable to NGC~2903 considering the relationship between $X_{\rm CO}$ and metallicity, 12 + log(O/H).
In general, $X_{\rm CO}$ decreases with the increase in metallicity because the CO abundance should be proportional to the carbon and oxygen abundances (e.g., \cite{arimoto1996}; \cite{boselli2002}).
In addition, it is reported that metallicity decreases with the galactocentric distance in NGC~2903 (e.g., \cite{dack2012}; \cite{pilyugin2014}).
These observational facts suggest a smaller $X_{\rm CO}$ by a factor of 1.5 -- 2 at the center than in the disk of NGC~2903, which yields a smaller gas mass, providing a higher SFE than the present one shown in table~3 and figure~11.
However, even if a higher SFE by a factor of 2 is adopted for the center, the global trend of the correlations shown in figure~11 does not change so much
because the original SFE at the center is already the highest in NGC~2903.
Therefore, we concluded that variations in the [$^{12}$CO]/[$^{13}$CO] abundance ratio and $X_{\rm CO}$ do $not$ affect the correlations of SFE with $n_{\rm H_2}$ and $T_{\rm K}$ in NGC~2903.

Note that the smaller $X_{\rm CO}$ corresponds to the larger $Z$($^{12}$CO) at the center of NGC~2903,
but the larger $dv/dr$ is also suggested because the typical velocity width at the center (250 -- 300 km s$^{-1}$) is
wider than those in other regions (150 -- 200 km s$^{-1}$) due to the rapid rotation of molecular gas near the galaxy center.
Thus we consider that $Z(^{12}$CO)/($dv/dr$) itself does not differ between the center and the disk in NGC~2903
even if the $Z(^{12}$CO) at the center is larger than that in the disk.

Finally, we examine the correlation coefficient for the least-square power-law fit $R^2$ between SFE and $n_{\rm H_2}$ and that between SFE and $T_{\rm K}$ shown in figure~11.
We found that the former is 0.50 and the latter is 0.08.
The significant correlation between SFE and $n_{\rm H_2}$ with $R^2$ of 0.50 suggests that molecular gas density governs the spatial variations in SFE.
This speculation is well consistent with earlier studies based on HCN emission (e.g., \cite{gao2004}; \cite{gao2007}; \cite{muraoka2009}; \cite{usero2015}).
In order to confirm whether such a relationship between SFE and $n_{\rm H_2}$ is applicable to other galaxies or not, we will perform further analysis toward other COMING sample galaxies,
considering variations in the [$^{12}$CO]/[$^{13}$CO] abundance ratio, $X_{\rm CO}$, and $Z(^{12}$CO)/($dv/dr$), in forthcoming papers.

\section{Summary}

We have performed the simultaneous mappings of $J=1-0$ emission of $^{12}$CO, $^{13}$CO, and C$^{18}$O molecules
toward the whole disk ($8' \times 5'$ or $20.8 \times 13.0$ kpc at the distance of 8.9 Mpc) of the nearby barred spiral galaxy NGC~2903
with the NRO 45-m telescope equipped with the FOREST at an effective angular resolution of $20''$ (or 870 pc).
A summary of this work is as follows.

\begin{enumerate}
\item
We detected $^{12}$CO($J=1-0$) emission over the disk of NGC~2903.
In addition, significant $^{13}$CO($J=1-0$) emission was found at the center and bar-ends, whereas we could not detect any significant C$^{18}$O($J=1-0$) emission.

\item
In order to improve the S/N of CO emission and to measure $R_{2-1/1-0}$ and $R_{13/12}$ with high significance,
we performed the stacking analysis for our $^{12}$CO($J=1-0$), $^{13}$CO($J=1-0$), and archival $^{12}$CO($J=2-1$) spectra with velocity-axis alignment
in nine representative regions (i.e., center, northern bar, southern bar, northern bar-end, southern bar-end, northern arm, southern arm, inter-arm, and outer-disk) of NGC~2903.
We successfully obtained the stacked CO spectra with highly improved S/N,
and thus we could confirm the significant $^{13}$CO($J=1-0$) emission for all the regions.

\item
We examined the averaged $R_{2-1/1-0}$ and $R_{13/12}$ for nine regions,
and found that the averaged $R_{2-1/1-0}$ shows the highest value of 0.92 at the center, and moderate or lower values of 0.6 -- 0.8 are observed in the disk.
However, the highest $R_{13/12}$ of 0.19 is observed not at the center but in the northern arm.
The central $R_{13/12}$ of 0.11 is similar to those in other regions (0.08 -- 0.13) except for the northern arm and outer-disk ($\sim$ 0.04).

\item
We determined $n_{\rm H_2}$ and $T_{\rm K}$ of molecular gas using $R_{2-1/1-0}$ and $R_{13/12}$ based on the LVG approximation.
Under the assumption of [$^{12}$CO]/[$^{13}$CO] abundance ratio of 50,
the derived $n_{\rm H_2}$ ranges from $\sim$1000 cm$^{-3}$ (in the bar, bar-ends, and spiral arms) to 3700 cm$^{-3}$ (at the center)
and the derived $T_{\rm K}$ ranges from 10 K (in the bar and spiral arms) to 30 K (at the center).

\item
We examined the dependence of SFE on $n_{\rm H_2}$ and $T_{\rm K}$ of molecular gas,
and found the positive correlation between SFE and $n_{\rm H_2}$ with the correlation coefficient for the least-square power-law fit $R^2$ of 0.50.
This suggests that molecular gas density governs the spatial variations in SFE.

\end{enumerate}

\vspace{0.5cm}
We thank the referee for invaluable comments, which significantly improved the manuscript.
We are indebted to the NRO staff for the commissioning and operation of the 45-m telescope and their continuous efforts to improve the performance of the instruments.
This work is based on observations at NRO, which is a branch of the National Astronomical Observatory of Japan, National Institutes of Natural Sciences.
This research has made use of the NASA/IPAC Extragalactic Database, which is operated by the Jet Propulsion Laboratory,
California Institute of Technology, under contract with the National Aeronautics and Space Administration.


\begin{figure}
  \begin{center}
    \includegraphics[width=8cm]{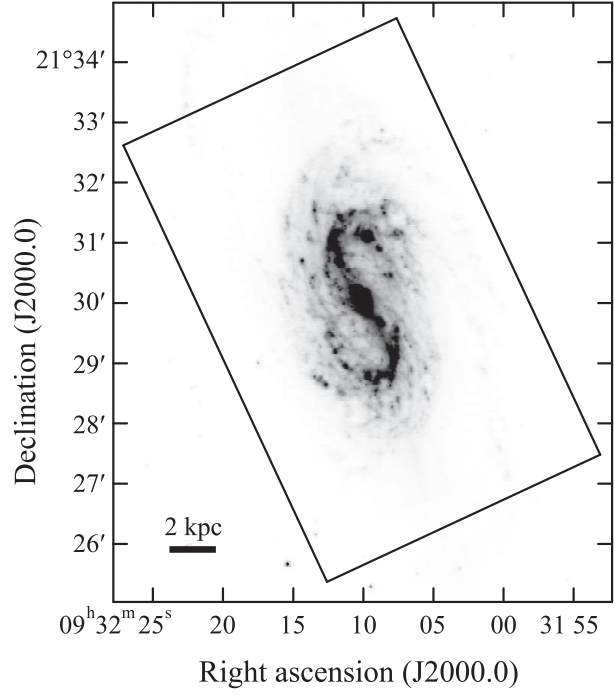}
  \end{center}
\caption{
Observed $8' \times 5'$ area of NGC~2903 with the FOREST mounted on the NRO 45-m telescope, which is indicated by a large square,
superposed on $Spitzer$/IRAC 8 $\micron$ image \citep{kennicutt2008}.
}
\label{fig:fig1}
\end{figure}

\begin{figure}
  \begin{center}
    \includegraphics[width=8cm]{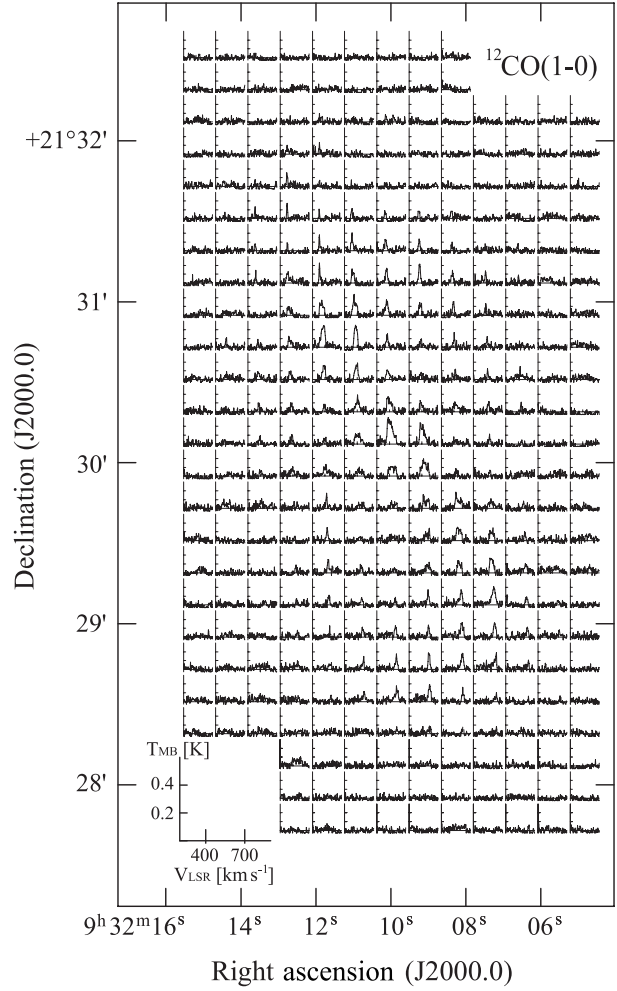}
  \end{center}
\caption{
Global spectra of $^{12}$CO($J=1-0$) emission in NGC~2903. The grid spacing was set to 12$''$ in order to display the spectrum in each pixel clearly.
The temperature scale of spectra is indicated by the small box inserted in the lower left corner.
}
\label{fig:fig2}
\end{figure}

\begin{figure}
  \begin{center}
    \includegraphics[width=8cm]{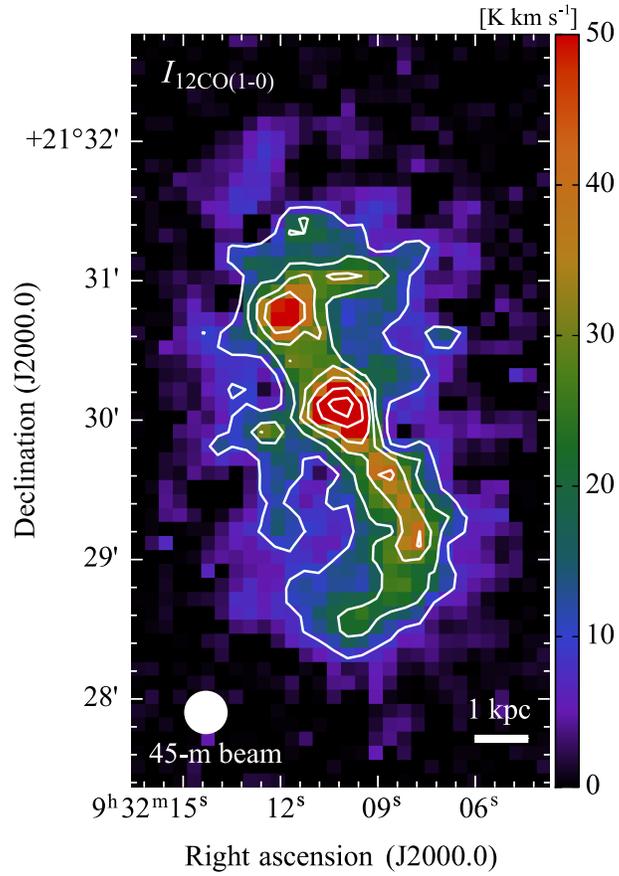}
  \end{center}
\caption{
A map of $I_{\rm 12CO(1-0)}$ of NGC~2903 obtained by COMING.
The contour levels are 10, 20, 30, 40, 60, and 80 K km s$^{-1}$.
}
\label{fig:fig3}
\end{figure}

\begin{figure}
  \begin{center}
    \includegraphics[width=8cm]{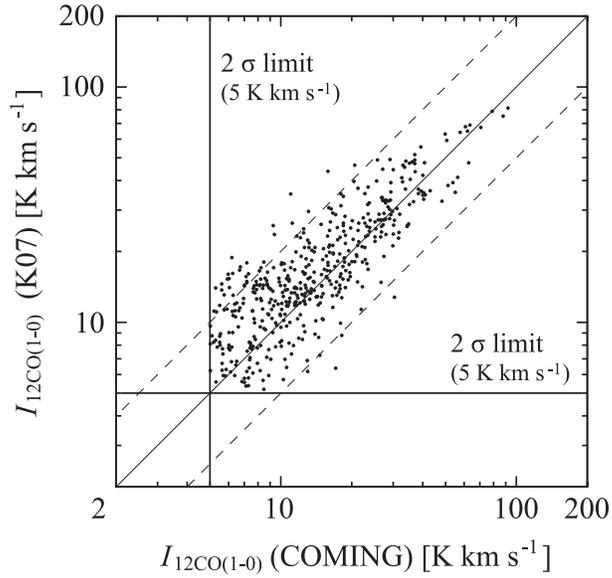}
  \end{center}
\caption{
A pixel-by-pixel comparison between $I_{\rm 12CO(1-0)}$ in NGC~2903 obtained by COMING and those obtained by K07.
The vertical and horizontal lines indicate the 2 $\sigma$ of $I_{\rm 12CO(1-0)}$ for COMING and K07 data, respectively.
The diagonal solid line indicates the ratio of $I_{\rm 12CO(1-0)}$ by COMING to those by K07 of unity, and the dashed lines indicate the ratio of 0.5 and 2.0.
Both $I_{\rm 12CO(1-0)}$ are well correlated with each other.
}
\label{fig:fig4}
\end{figure}

\begin{figure}
  \begin{center}
    \includegraphics[width=17cm]{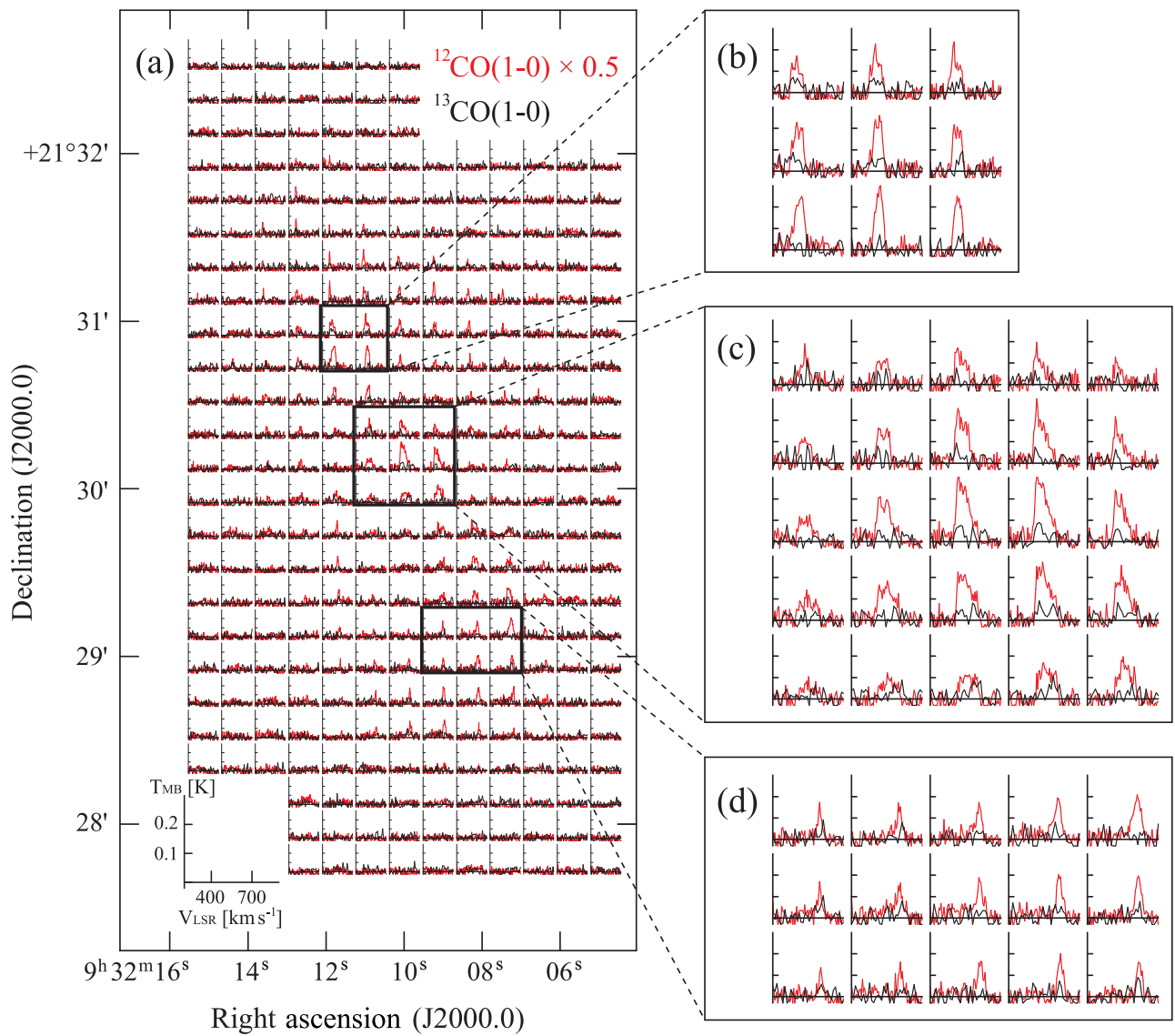}
  \end{center}
\caption{
(a) Global spectra of $^{13}$CO($J=1-0$) emission in NGC~2903.
For comparison, spectra of $^{12}$CO($J=1-0$) emission multiplied by 0.5 are overlaid in red line.
As well as figure~2, the grid spacing was set to 12$''$.
The temperature scale of spectra is indicated by the small box inserted in the lower left corner.
(b) Magnified spectra with the grid spacing of 6$''$ at the northern bar-end.
(c) Same as (b), but at the center.
(d) Same as (b), but at the southern bar-end.
}
\label{fig:fig5}
\end{figure}

\begin{figure}
  \begin{center}
    \includegraphics[width=8cm]{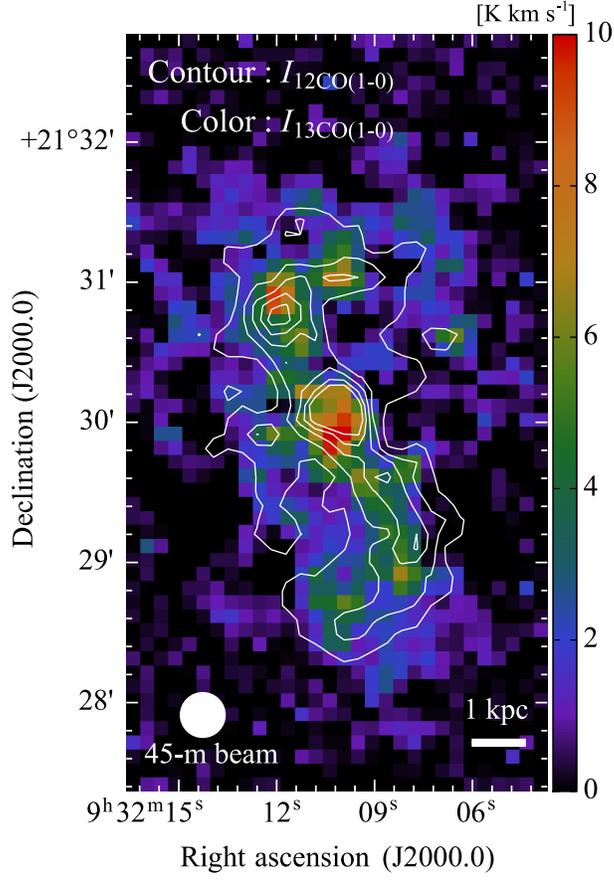}
  \end{center}
\caption{
A color map of $I_{\rm 13CO(1-0)}$ superposed on the contour map of $I_{\rm 12CO(1-0)}$ of NGC~2903.
The contour levels of $I_{\rm 12CO(1-0)}$ are the same as figure~3.
}
\label{fig:fig6}
\end{figure}

\begin{figure}
  \begin{center}
    \includegraphics[width=17cm]{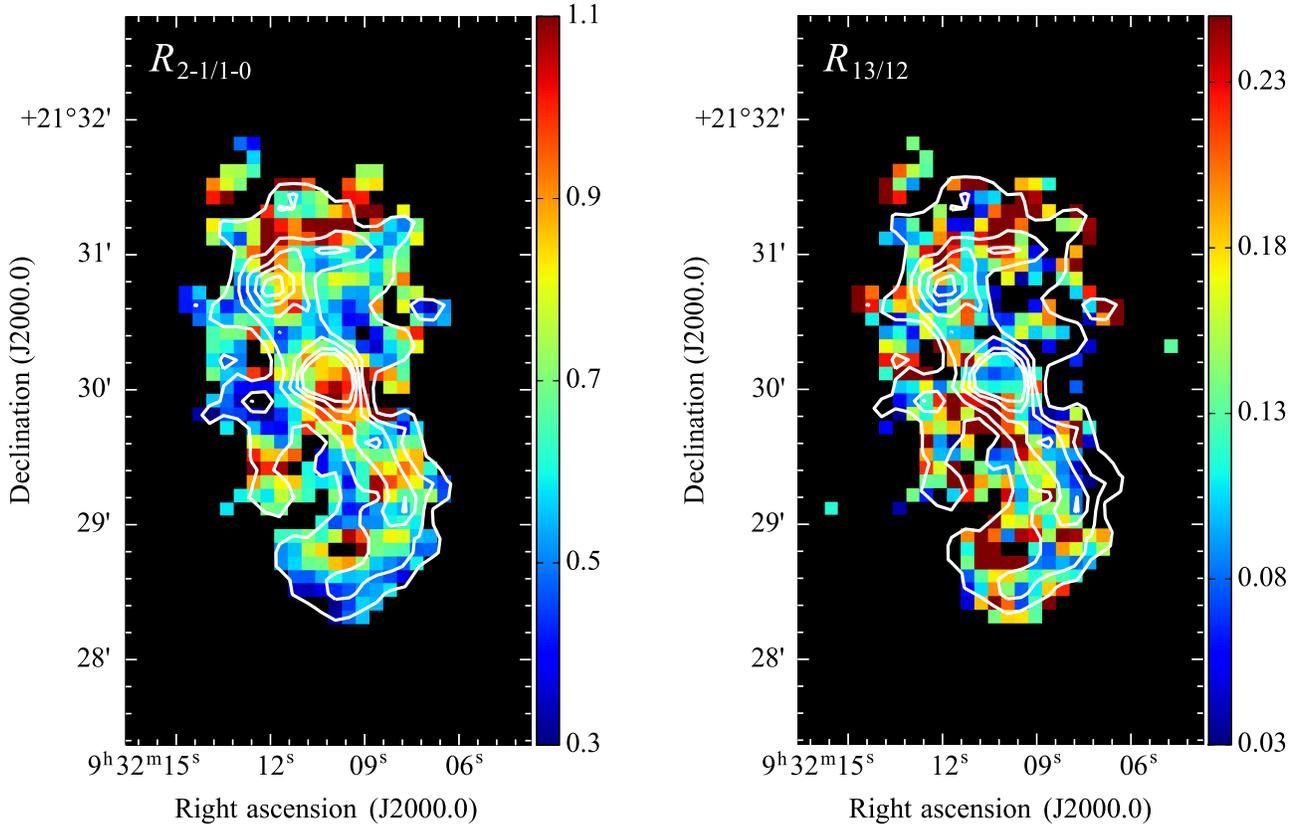}
  \end{center}
\caption{
Spatial distributions of $R_{2-1/1-0}$ (left) and $R_{13/12}$ (right) superposed on the contour map of $I_{\rm 12CO(1-0)}$ of NGC~2903.
The contour levels of $I_{\rm 12CO(1-0)}$ are the same as figure~3.
There are some local peaks of $R_{2-1/1-0}$ ($\sim$ 1.0) near the center and at the downstream side of the northern spiral arm,
whereas lower $R_{2-1/1-0}$ ($\sim$ 0.5 -- 0.6) was observed in the bar.
The spatial distribution of $R_{13/12}$ is unclear due to poor S/N of $^{13}$CO($J=1-0$) emission.
}
\label{fig:fig7}
\end{figure}

\begin{figure}
  \begin{center}
    \includegraphics[width=17cm]{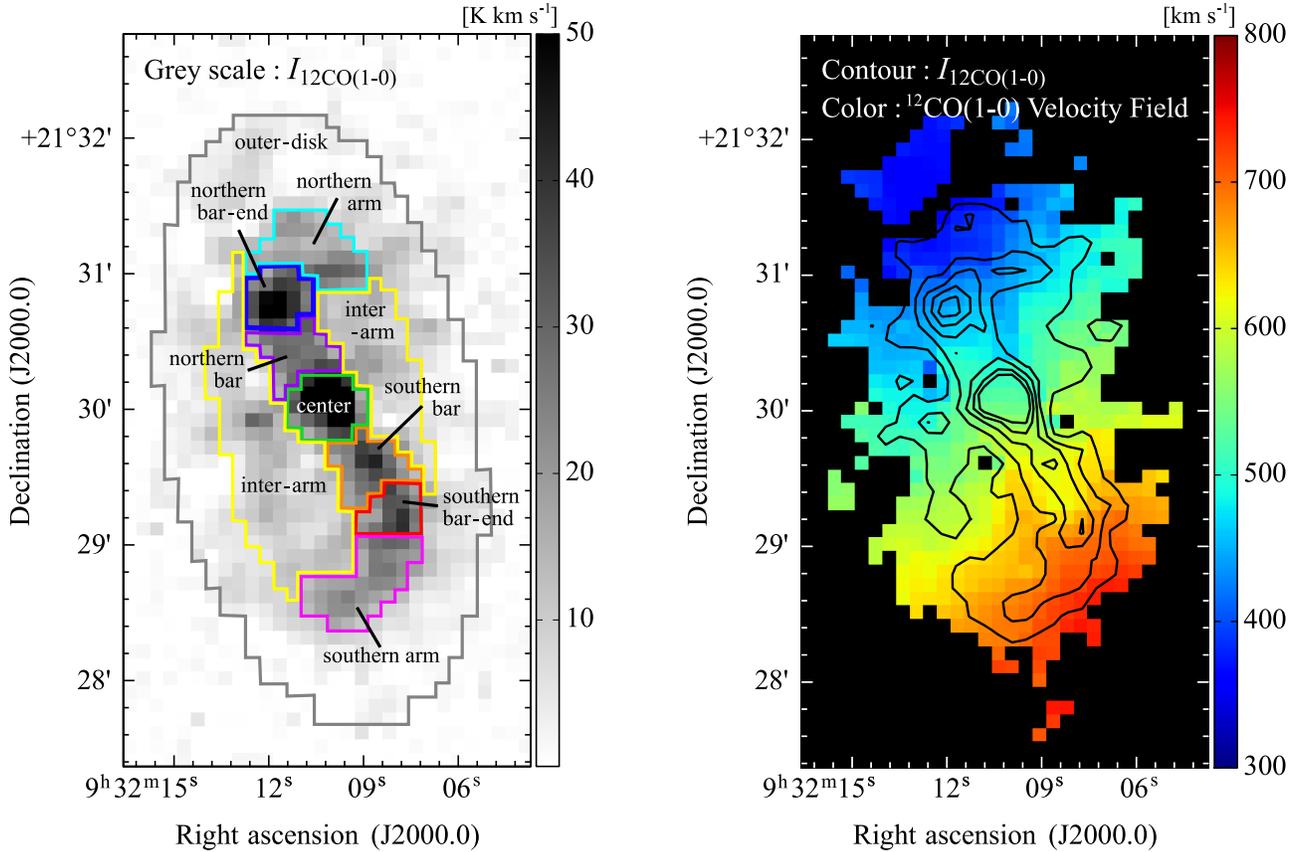}
  \end{center}
\caption{
(left) Separation of representative regions for stacking analysis superposed on the grey-scale map of $I_{\rm 12CO(1-0)}$ of NGC~2903.
The green frame indicates the center, the purple and orange indicate the northern and southern bars,
the blue and red indicate the northern and southern bar-ends,
the cyan and magenta indicate the northern and southern arms,
the yellow indicates the inter-arm, and the grey indicates the outer-disk.
(right) Intensity-weighted mean velocity field calculated from $^{12}$CO($J=1-0$) data superposed on the contour map of $I_{\rm 12CO(1-0)}$.
The contour levels of $I_{\rm 12CO(1-0)}$ are the same as figure~3.
}
\label{fig:fig8}
\end{figure}

\begin{figure}
  \begin{center}
    \includegraphics[width=17cm]{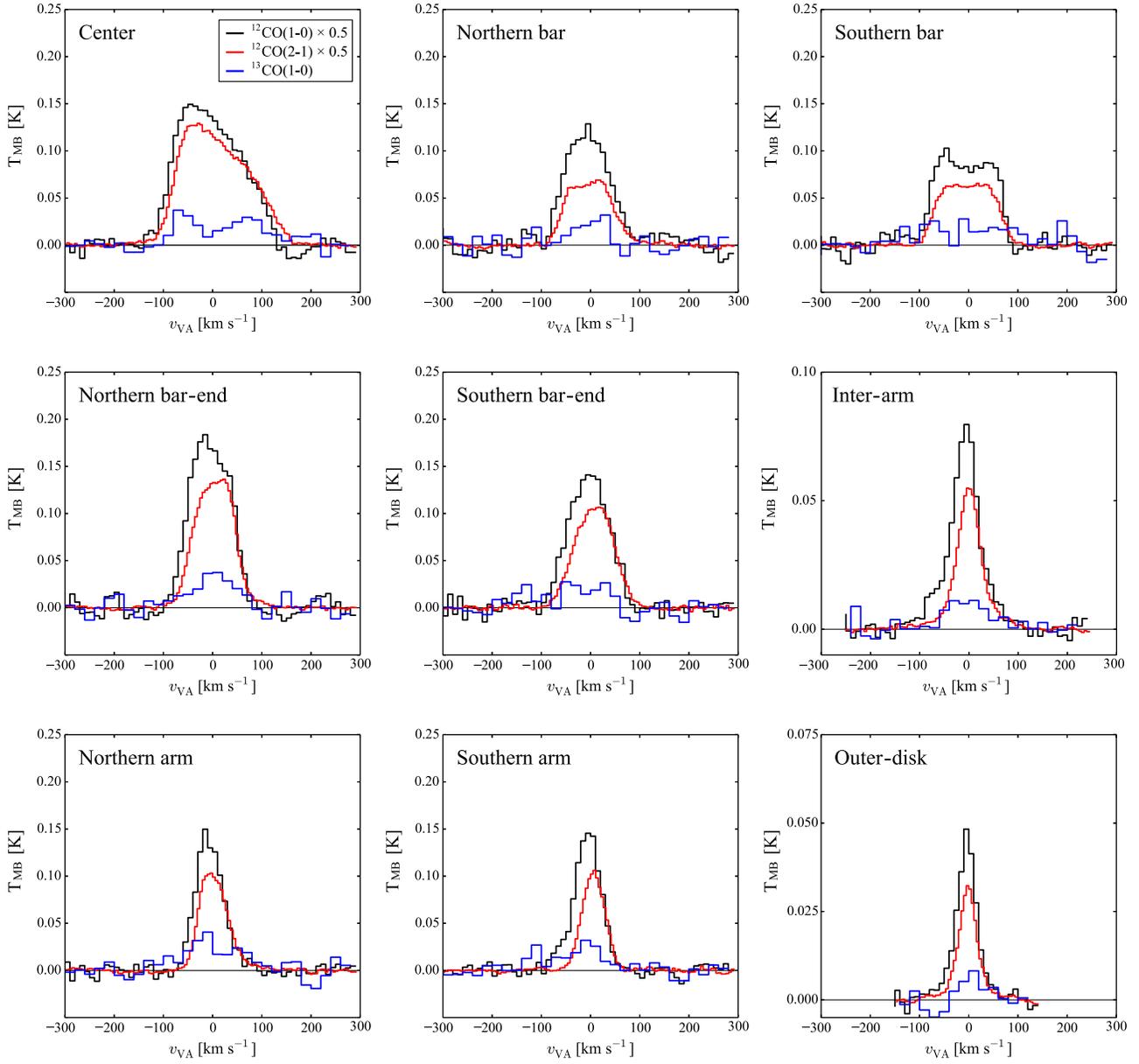}
  \end{center}
\caption{
Stacked CO spectra for each region in NGC~2903.
The black line indicates $^{12}$CO($J=1-0$) emission multiplied by 0.5, the red indicates $^{12}$CO($J=2-1$) emission multiplied by 0.5, and the blue indicates  $^{13}$CO($J=1-0$) emission.
The S/N of each CO emission is dramatically improved, and thus a significant $^{13}$CO($J=1-0$) emission is confirmed for all the regions.
}
\label{fig:fig9}
\end{figure}

\begin{figure}
  \begin{center}
    \includegraphics[width=17cm]{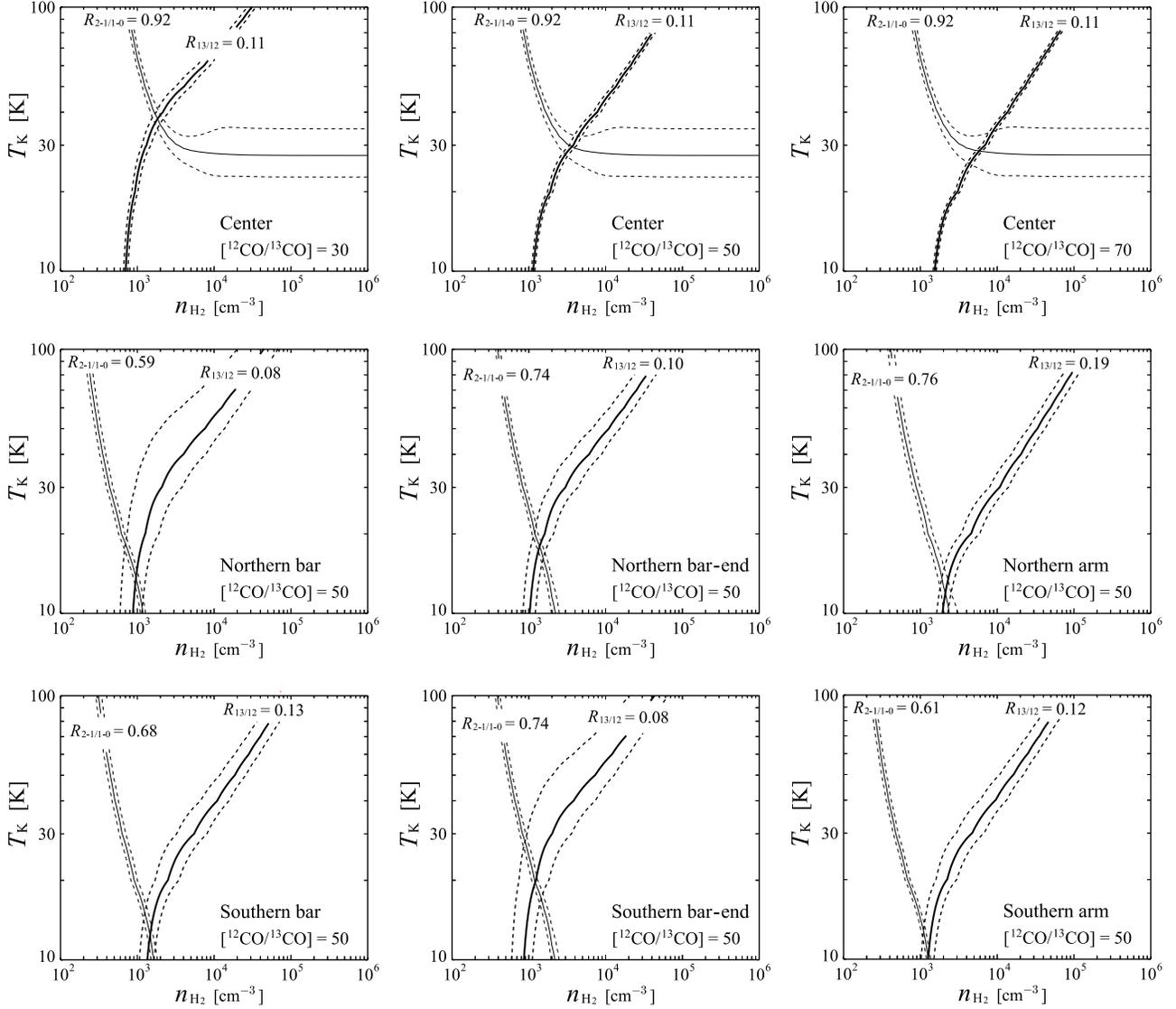}
  \end{center}
\caption{
Curves of constant $R_{2-1/1-0}$ (thin line) and $R_{13/12}$ (thick line) as functions of molecular gas density $n_{\rm H_2}$ and kinetic temperature $T_{\rm K}$.
$^{12}$CO fractional abundance per unit velocity gradient $Z(^{12}$CO)/($dv/dr$) was assumed to be $1.0 \times 10^{-5}$ (km s$^{-1}$ pc$^{-1}$)$^{-1}$.
The [$^{12}$CO]/[$^{13}$CO] abundance ratio was assumed to be a fixed value of 50 for the bar, bar-ends, and spiral arms,
but three different [$^{12}$CO]/[$^{13}$CO] abundance ratios of 30, 50, and 70 were assumed for the center. 
Dashed lines indicate $\pm1 \sigma$ error of each line ratio.
}
\label{fig:fig10}
\end{figure}

\begin{figure}
  \begin{center}
    \includegraphics[width=17cm]{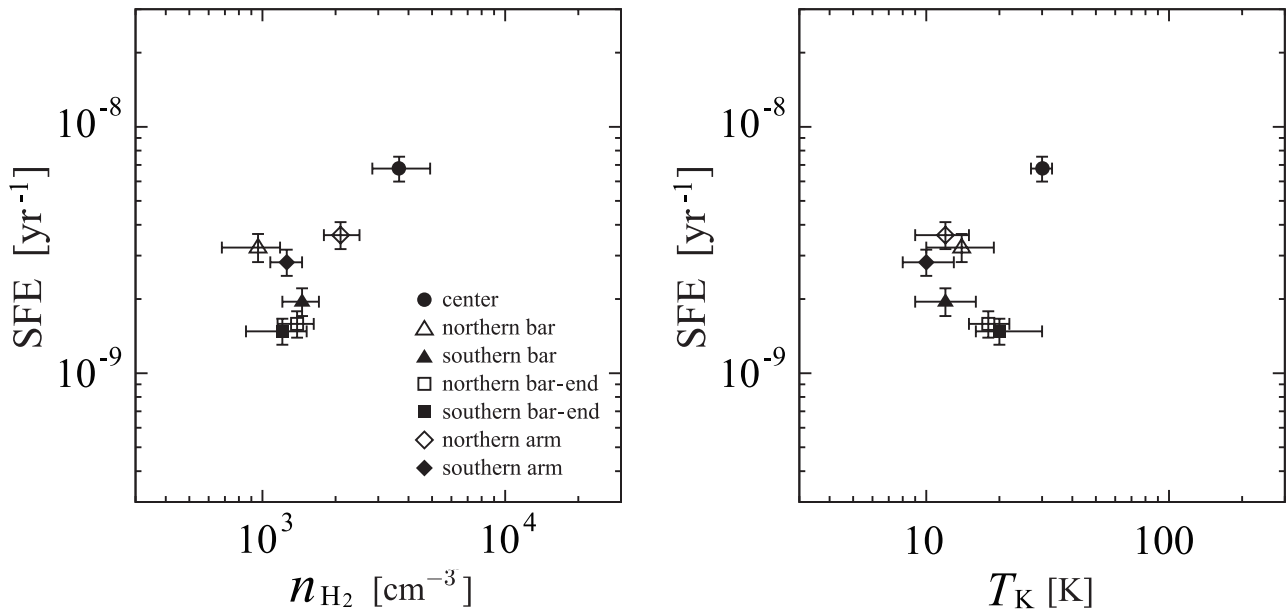}
  \end{center}
\caption{
Correlation between SFE and $n_{\rm H_2}$ (left) and that between SFE and $T_{\rm K}$ (right) in each region of NGC~2903.
}
\label{fig:fig11}
\end{figure}


\begin{table}
\begin{center}
Table~1.\hspace{4pt}General parameters of NGC~2903\\[1mm]
\begin{tabular}{lc}
\hline \hline \\[-6mm]
Morphological type$^{\rm a}$ & SAB(rs)bc \\
Map center$^{\rm b}$: & \\[-1mm]
\,\,\,\, Right Ascension (J2000.0) & \timeform{09h32m10s.1} \\[-1mm]
\,\,\,\, Declination (J2000.0) & \timeform{21D30'03''.0} \\
Distance$^{\rm c}$ & 8.9 Mpc \\
Linear scale & 43 pc arcsec$^{-1}$  \\
Inclination$^{\rm d}$ & \timeform{65D} \\
\hline \\[-2mm]
\end{tabular}\\
{\footnotesize
$^{\rm a}$Morphological type from RC3.
$^{\rm b}$Map center from \citet{jarrett2003}. \\
$^{\rm c}$Adopted distance from \citet{drozdovsky2000}.
$^{\rm d}$Inclination from \citet{deblok2008}.
}
\end{center}
\end{table}


\begin{table}
\begin{center}
Table~2.\hspace{4pt}Averaged line intensity and line ratios based on the stacking analysis\\[1mm]
\begin{tabular}{lccccc}
\hline \hline \\[-6mm]
  & $I_{\rm 12CO(1-0)}$ & $I_{\rm 12CO(2-1)}$ & $I_{\rm 13CO(1-0)}$ & $R_{2-1/1-0}$ & $R_{13/12}$ \\[-2mm]
  & [K km s$^{-1}$] & [K km s$^{-1}$] & [K km s$^{-1}$] & & \\
\hline \\[-6mm]
Center           & $46.7\pm0.7$ & $43.0\pm0.1$ & $5.1\pm0.2$ & $0.92\pm0.02$ & $0.11\pm0.01$ \\
Northern bar     & $26.4\pm0.8$ & $15.6\pm0.1$ & $2.2\pm0.6$ & $0.59\pm0.02$ & $0.08\pm0.02$ \\
Southern bar     & $26.6\pm0.8$ & $18.1\pm0.1$ & $3.5\pm0.7$ & $0.68\pm0.02$ & $0.13\pm0.03$ \\
Northern bar-end & $36.8\pm0.8$ & $27.3\pm0.1$ & $3.6\pm0.6$ & $0.74\pm0.02$ & $0.10\pm0.02$ \\
Southern bar-end & $28.6\pm0.6$ & $21.2\pm0.1$ & $2.4\pm0.6$ & $0.74\pm0.02$ & $0.08\pm0.02$ \\
Northern arm     & $19.8\pm0.5$ & $15.0\pm0.1$ & $3.8\pm0.6$ & $0.76\pm0.02$ & $0.19\pm0.02$ \\
Southern arm     & $19.6\pm0.5$ & $11.9\pm0.1$ & $2.4\pm0.4$ & $0.61\pm0.02$ & $0.12\pm0.02$ \\
Inter-arm        & $11.8\pm0.3$ & $ 7.2\pm0.1$ & $1.2\pm0.3$ & $0.61\pm0.02$ & $0.10\pm0.02$ \\
Outer-disk       & $ 5.0\pm0.2$ & $ 3.2\pm0.1$ & $0.2\pm0.1$ & $0.64\pm0.02$ & $0.04\pm0.02$ \\
\hline \\[-2mm]
\end{tabular}\\
{\footnotesize
}
\end{center}
\end{table}


\begin{table}
\begin{center}
Table~3.\hspace{4pt}Derived $n_{\rm H_2}$ and $T_{\rm K}$, and SFE in each region of NGC~2903\\[1mm]
\begin{tabular}{lcccc}
\hline \hline \\[-6mm]
  region         & [$^{12}$CO]/[$^{13}$CO]  & $n_{\rm H_2}$            & $T_{\rm K}$     & SFE \\[-2mm]
                 &                          & [$10^3$ cm$^{-3}$]       & [K]             & [$10^9$ yr$^{-1}$]\\
\hline \\[-6mm]
Center           & 50                       & $3.7^{+1.2}_{-0.9}$      & $30\pm3$        & $6.8\pm0.8$ \\
Northern bar     & 50                       & $0.96^{+0.22}_{-0.28}$   & $14^{+5}_{-4}$  & $3.2\pm0.4$ \\
Southern bar     & 50                       & $1.5^{+0.2}_{-0.3}$      & $12^{+4}_{-3}$  & $2.0\pm0.3$ \\
Northern bar-end & 50                       & $1.4\pm0.2$              & $18^{+4}_{-3}$  & $1.6\pm0.2$ \\
Southern bar-end & 50                       & $1.2\pm0.3$              & $20^{+10}_{-4}$ & $1.5\pm0.2$ \\
Northern arm     & 50                       & $2.1^{+0.4}_{-0.3}$      & $12\pm3$        & $3.6\pm0.5$ \\
Southern arm     & 50                       & $1.3\pm0.2$              & $10^{+3}_{-2}$  & $2.8\pm0.4$ \\[2mm]
Center           & 30                       & $1.8^{+0.4}_{-0.2}$      & $38\pm3$        & $6.8\pm0.8$ \\
Center           & 70                       & $5.9^{+3.9}_{-1.8}$      & $29^{+5}_{-4}$  & $6.8\pm0.8$ \\
\hline \\[-2mm]
\end{tabular}\\
{\footnotesize
[$^{12}$CO]/[$^{13}$CO] means the assumed [$^{12}$CO]/[$^{13}$CO] abundance ratio for the LVG calculation.
}
\end{center}
\end{table}


\end{document}